# Quadrupole terms in the Maxwell equations: Born energy, partial molar volume and entropy of ions. Debye-Hückel theory in a quadrupolarizable medium

**Radomir I. Slavchov,**[*,a] **Tzanko I. Ivanov**[b]

[a] *Department of Physical Chemistry, Faculty of Chemistry and Pharmacy, Sofia University, 1164 Sofia, Bulgaria*
[b] *Theoretical Physics Department, Faculty of Physics, Sofia University, 1164 Sofia, Bulgaria*
* E-mail:fhri@chem.uni-sofia.bg

A new equation of state relating the macroscopic quadrupole moment density $\mathbf{Q}$ to the gradient of the field $\nabla \mathbf{E}$ in an isotropic fluid is derived: $\mathbf{Q} = \alpha_Q(\nabla \mathbf{E} - \mathbf{U}\nabla\cdot\mathbf{E}/3)$, where the quadrupolarizability $\alpha_Q$ is proportional to the squared molecular quadrupole moment. Using this equation of state, a generalized expression for the Born energy of an ion dissolved in quadrupolar solvent is obtained. It turns out that the potential and the energy of a point charge in a quadrupolar medium are finite. From the obtained Born energy, the partial molar volume and the partial molar entropy of a dissolved ion follow. Both are compared to experimental data for a large number of simple ions in aqueous solutions. From the comparison the value of the quadrupolar length $L_Q$ is determined, $L_Q = (\alpha_Q/3\varepsilon)^{1/2} = $ 1-2 Å. Further, the extended Debye-Hückel model is generalized to ions in a quadrupolar solvent. If quadrupole terms are allowed in the macroscopic Coulomb law, they result in suppression of the gradient of the electric field. In result, the electric double layer is slightly expanded. The activity coefficients obtained within this model involve three characteristic lengths: Debye length, ion radius and quadrupolar length $L_Q$. Comparison to experimental data shows that minimal distance between ions is equal to the sum of their bare ion radii; the concept for ion hydration as an obstacle for ions to come into contact is not needed for the understanding of the experimental data.

## 1. Introduction

The macroscopic Poisson equation of electrostatics combines the static macroscopic Coulomb and Ampere laws,

$$\nabla \cdot \mathbf{D} = \rho(\phi); \quad (1)$$

$$\mathbf{E} = -\nabla\phi, \quad (2)$$

with a linear dependence of the electric displacement field $\mathbf{D}$ on the electric field intensity $\mathbf{E}$

$$\mathbf{D} \equiv \varepsilon_0 \mathbf{E} + \mathbf{P} = \varepsilon_0 \mathbf{E} + \alpha_P \mathbf{E} = \varepsilon \mathbf{E}. \quad (3)$$

Here $\rho$ is the free charge number density; $\phi$ is the electrostatic potential; $\varepsilon \equiv \varepsilon_0 + \alpha_P = \varepsilon_0 \varepsilon_r$ is the absolute dielectric permittivity, $\varepsilon_0$ is the vacuum permittivity, $\varepsilon_r$ is the relative permittivity of the medium, $\alpha_P$ is the macroscopic polarizability of the medium. For a homogeneous medium ($\nabla\varepsilon = 0$) the Poisson equation for $\phi$ follows from Eqs (1)-(3):

$$-\varepsilon\nabla^2\phi = \rho(\phi). \quad (4)$$

For conducting media, one must provide also an equation of state for the dependence $\rho(\phi)$. A common assumption is that the charges are distributed according to the Boltzmann distribution

$$\rho(\phi) = \sum e_i C_i \exp(-e_i\phi/T), \quad (5)$$

where $e_i = eZ_i$ is the absolute charge of the $i^{th}$ ion, $e$ is the electron charge, $Z_i$ is the relative ionic charge, $C_i = \nu_i C_{el}$ is the local concentration of the $i^{th}$ ion, $\nu_i$ stands for its stoichiometric number, $C_{el}$ is the electrolyte concentration, $T[\mathrm{J}] = k_B T[\mathrm{K}]$ is the thermodynamic temperature. Inserting Eq (5) into Eq (4), one obtains the Poisson-Boltzmann equation, widely used in physical chemistry and colloid science. Numerous basic concepts such as Debye-Hückel double layer[1,2], Gouy model for charged interface[3,4], Davies adsorption model for ionic surfactant adsorption[5,6], the electrostatic disjoining pressure in DLVO theory[7-9], electrokinetic $\zeta$-potential[10], etc., are merely a consequence of Eqs (4)-(5).

It has been early recognized that both Poisson and Boltzmann equations (4)-(5) have severe limitations. The derivation of Eq (4) involves a multipole expansion of the local potential up to dipole terms, i.e., it neglects the quadrupole moment density[11-13]. Eq (3) is strictly valid for linear media[14]. The Boltzmann distribution (5) is only a first approximation valid for ideal solution[15-17]; other "external" potentials except $e_i\phi$ often arise[18-20]. In order to make Eqs (3)-(5) applicable to real systems, numerous corrections have been proposed, to point a few: **(i)** Corrections to the Boltzmann distribution (5) by introduction of various additional interaction potentials, either for ion-ion non-electrostatic interaction[15-16] or various ion-surface interactions[18-20]; **(ii)** Corrections for the macroscopic nature of the equation, involving explicit molecular treatment of the first neighbor interactions[21,14,22] or other discreteness effects[23-24]; **(iii)** Correction for the dielectric saturation, i.e., the dependence of $\varepsilon$ on the electric field intensity[25,26,14] $E$; **(iv)** corrections related to the inhomogeneity of the medium ($\nabla\varepsilon \neq 0$; e.g., Refs. 27,28); **(v)** correlation effects[2,17] and non-local electrostatic effects[29-32], etc. Every major correction of Eqs (4)-(5) have been an impetus for reconsideration of the basic concepts following from the Poisson-Boltzmann equation.

While most studies in physical chemistry criticized mainly the Boltzmann part of Poisson-Boltzmann equation, several studies of optical phenomena[33-36] attacked the Poisson part of it. It was demonstrated that the quadrupolar terms in the macroscopic Coulomb law (1) become quite significant in cases where high gradients of $\mathbf{E}$ are present. In such cases, quadrupolar term in the displacement field $\mathbf{D}$ need to be introduced[36]:

$$\mathbf{D} = \varepsilon_0\mathbf{E} + \mathbf{P} - \tfrac{1}{2}\nabla\cdot\mathbf{Q}. \quad (6)$$

Here, $\mathbf{Q}$ is the macroscopic density of the quadrupole moment tensor. Note that the coefficient in front of $\nabla\cdot\mathbf{Q}$ depends on the choice of definition of quadrupole moment (for convenience, a derivation of Eq (6) and the definitions of the involved quantities are given in Supplementary information A[†]). The substitution of

Eq (6) into Eq (1) yields a generalization of the Poisson equation (4),

$$\nabla \cdot \left(\varepsilon \boldsymbol{E} - \frac{1}{2}\nabla \cdot \boldsymbol{Q}\right) = \rho, \quad (7)$$

which opens a vast field for analysis of the effect of the quadrupole moments of the molecules composing a medium on the properties of charged particles in such medium. It have been recently demonstrated that quadrupole terms in $\boldsymbol{D}$ can play a role in solvent-solute interaction[12,13]. The correction for $\boldsymbol{Q}$ will be important if the solvent molecules possess large quadrupole moment – such is the case of water[37] and many others, including "non-polar" media of low dipole moment but high quadrupole moment such as liquid $CO_2$, fluorocarbons etc.[13,38].

The purpose of our study is to analyze the consequences of the new term in Maxwell equation for several basic problems of physical chemistry of electrolyte solutions and colloid chemistry. Eq (7) is largely unknown to physical chemists and virtually has never been used in colloid science. There are three reasons for this negligence. First, Eq (7) is useless without an equation of state for $\boldsymbol{Q}$. There are several existing studies of this constitutive relation[39-42,12,13,36] but all are scarcely analyzed. Therefore, in Section 2, we will derive a new equation of state as simple as possible, showing that $\boldsymbol{Q}$ is a linear function of $\nabla \boldsymbol{E} - \boldsymbol{U}\nabla\cdot\boldsymbol{E}/3$, with a scalar coefficient of proportionality – the *quadrupolarizability*[33] $\alpha_Q$ (here $\boldsymbol{U}$ is the unit tensor). The second reason for Eq (7) is unknown in the colloid field is that it is a fourth-order equation with respect to $\phi$, and requires the use of new boundary conditions. Seemingly, these new conditions have been derived only recently[39,40]. We will review this problem in Section 3. Finally, the third obstacle to use Eq (7) is that it involves a new parameter of unknown value – $\alpha_Q$. We will give in this paper both theoretical estimation and values determined from 3 independent sets of experimental data for ions in water.

In Sections 4 and 5, Eq (7) is used to reinvestigate the most basic concepts in the physical chemistry of electrolyte solutions – Born energy and Debye-Hückel diffuse ionic atmosphere. It will be shown that the "correction" $\boldsymbol{Q}$ in Eq (7) in fact leads to results which have no counterpart in the frame of Poisson equation (4), notably, finite electrostatic potential and energy of a point charge in quadrupolarizable medium (similar result was obtained in Ref. 12). In Sections 4 and 5, we compare our results for both the Born energy and the Debye-Hückel diffuse atmosphere in quadrupolar medium to experimental data, which allows us to determine the value of quadrupolarizability $\alpha_Q$ of water.

## 2. Equations of state for the quadrupole moment density

The problem for the constitutive relation between $\boldsymbol{Q}$ and the field gradient $\nabla \boldsymbol{E}$ has been addressed several times[33,36,39-42,12,13]. Using as a starting point the approach of Jeon and Kim[13], we will be able to obtain a new simple equation of state which relates $\boldsymbol{Q}$ to the field gradient $\nabla \boldsymbol{E}$ and the molecular properties of the solvent.

Consider an ideal gas consisting of molecules possessing a solid quadrupole moment tensor $\boldsymbol{q}_0$ (for the sake of simplicity, the molecule is assumed non-polarizable and with no dipole moment). Since $\boldsymbol{q}_0$ is symmetrical and traceless, by a suitable choice of the coordinate system it can be diagonalized[43] and in the general case, its diagonal form is:

$$\boldsymbol{q}_0 = \begin{pmatrix} q_{xx} & 0 & 0 \\ 0 & q_{yy} & 0 \\ 0 & 0 & q_{zz} \end{pmatrix} - \frac{q_{xx}+q_{yy}+q_{zz}}{3}\boldsymbol{U}. \quad (8)$$

The term $(q_{xx}+q_{yy}+q_{zz})/3\boldsymbol{U}$ ensures that the trace of $\boldsymbol{q}_0$ is 0 and can be added because the field created by a quadrupole in vacuum is invariant with respect to the operation of exchanging the quadrupole strength $\boldsymbol{q}_0$ with $\boldsymbol{q}_0 + X\boldsymbol{U}$ where $X$ is any scalar[11,44]. The molecule is constantly rotating. An arbitrary rotation changes the quadrupole moment tensor from $\boldsymbol{q}_0$ to $\boldsymbol{q}$. For a rotation at arbitrary Eulerian angles $\phi$, $\psi$ and $\theta$, the Euler matrix $\boldsymbol{E}^{\phi\psi\theta}$ is given by:

$$\boldsymbol{E}^{\varphi\psi\theta} = \begin{pmatrix} \cos\varphi\cos\psi - & \sin\varphi\cos\psi + & \\ -\sin\varphi\sin\psi\cos\theta & +\cos\varphi\sin\psi\cos\theta & \sin\psi\sin\theta \\ -\cos\varphi\sin\psi - & -\sin\varphi\sin\psi + & \\ -\sin\varphi\cos\psi\cos\theta & +\cos\varphi\cos\psi\cos\theta & \cos\psi\sin\theta \\ \sin\varphi\sin\theta & -\cos\varphi\sin\theta & \cos\theta \end{pmatrix}. \quad (9)$$

The relation between the tensor $\boldsymbol{q}$ for a randomly oriented molecule and the tensor $\boldsymbol{q}_0$ is

$$q_{ij}(\varphi,\psi,\theta) = E^{\varphi\psi\theta}_{ik} E^{\varphi\psi\theta}_{jl} q_{0kl}. \quad (10)$$

In the absence of a gradient of the electric field the average value of $\boldsymbol{q}$ is 0. In an external electric field gradient $\nabla\boldsymbol{E}$, the molecule tends to orientate itself in order to minimize its electric energy, given by the expression (*Eq 4.17* of Jackson[11]):

$$u_{el} = -\frac{1}{2}\boldsymbol{q}:\nabla\boldsymbol{E}. \quad (11)$$

The orientation of the molecule must follow the Boltzmann distribution which can be linearized in the case of $u_{el}/T \ll 1$:

$$\rho_{\phi\psi\theta} = c_n \exp(-u_{el}/T) \approx c_n(1 - u_{el}/T). \quad (12)$$

Here, $c_n$ is a normalizing coefficient calculated as

$$c_n = 1/\int_0^{2\pi}\int_0^{2\pi}\int_0^{\pi}(1-u_{el}/T)\sin\theta\,d\theta\,d\varphi\,d\psi = 1/8\pi^2. \quad (13)$$

The average quadrupole moment $\bar{\boldsymbol{q}}$ of a molecule can be calculated directly using Eqs (8)-(13):

$$\bar{\boldsymbol{q}} = \int_0^{2\pi}\int_0^{2\pi}\int_0^{\pi}\boldsymbol{q}\rho_{\phi\psi\theta}\sin\theta\,d\theta\,d\varphi\,d\psi = \alpha_q(\nabla\boldsymbol{E} - \boldsymbol{U}\nabla\cdot\boldsymbol{E}/3). \quad (14)$$

Here, the molecular quadrupolarizability $\alpha_q$ was introduced, related to the diagonal components of $\boldsymbol{q}_0$ as follows:

$$\alpha_q = \boldsymbol{q}_0:\boldsymbol{q}_0/10T =$$
$$= (q_{xx}^2 + q_{yy}^2 + q_{zz}^2 - q_{xx}q_{yy} - q_{xx}q_{zz} - q_{yy}q_{zz})/15T. \quad (15)$$

Eq (15) was obtained e.g. in Ref. 13.

The macroscopic density $\boldsymbol{Q}$ of the quadrupole moment in a gas acted upon by a field gradient $\nabla\boldsymbol{E}$ is the gas concentration $C$ times $\bar{\boldsymbol{q}}$, Eq (14):

$$\boldsymbol{Q} = \alpha_Q(\nabla\boldsymbol{E} - \boldsymbol{U}\nabla\cdot\boldsymbol{E}/3). \quad (16)$$

Here, the macroscopic quadrupolarizability $\alpha_Q$ is given by

$$\alpha_Q = C\alpha_q = C\boldsymbol{q}_0:\boldsymbol{q}_0/10T. \quad (17)$$

Our constitutive relation Eq (16) is a direct consequence of the general form (8) of the molecular solid quadrupole and the linearized Boltzmann distribution (12). Note that according to Eq (16) $\boldsymbol{Q}$ is traceless[11], in contrast to *Eq 2.25* of Jeon and Kim[13]. Refs. 42 and 45 contain some discussion in favor of their choice. However, in Supplementary information A[†], we present arguments that the use of tensor $\mathbf{Q}$ with non-zero trace is incompatible with Eqs (6)-(7). *Eq 2.4* of Chitanvis[12] postulates an equation of state in which only the $\mathbf{U}\nabla\cdot\mathbf{E}$ term of our Eq (16) is present, i.e., according to him, $\mathbf{Q}$ has only diagonal elements and a non-zero trace.

In general, $\mathbf{Q}$ depends not only on the field gradient but also on the field $\mathbf{E}$ itself, and, on the other hand, electric field gradient $\nabla\mathbf{E}$ can induce non-zero dipole moment[39,36]. For an ideal gas of solid dipoles within the linear approximation for $\rho_{\phi\psi\theta}$, this is not the case. This can be shown by a direct calculation analogous to the derivation of Eq (16): if the molecule has dipole moment $\mathbf{p}_0$ and quadrupole moment $\mathbf{q}_0$, then in external field $\mathbf{E}$ and field gradient $\nabla\mathbf{E}$ its energy is[11]:

$$u_{el} = -\mathbf{p}\cdot\mathbf{E} - \frac{1}{2}\mathbf{q}:\nabla\mathbf{E} . \qquad (18)$$

Using this expression instead of Eq (11), one can calculate the average dipole and quadrupole moments. This calculation yields for $\mathbf{Q}$ again Eq (16), because the terms proportional to $\mathbf{E}$ cancel each other. Calculation of the macroscopic polarization $\mathbf{P}$ gives the classical result[14]:

$$\mathbf{P} = \alpha_P \mathbf{E}, \quad \alpha_P = C\alpha_p = Cp_0^2/3T , \qquad (19)$$

where $\alpha_p$ and $\alpha_P$ are the molecular and the macroscopic polarizabilities.

The derivation above is strictly valid for a gas of solid multipoles. It can be readily generalized to include molecular polarizabilities and quadrupolarizabilities[13]. This yields instead of Eq (17) the expression

$$\alpha_q = \bar{\alpha}_{q0} + \mathbf{q}_0 : \mathbf{q}_0 /10T , \qquad (20)$$

where $\bar{\alpha}_{q0}$ is the average intrinsic (atomic + electronic) molecular quadrupolarizability (*Eq 4.5* of Jeon and Kim[13]). Eq (20) can be compared to the well-known formula for the polarizability[14]

$$\alpha_p = \bar{\alpha}_{p0} + p_0^2/3T , \qquad (21)$$

where $\bar{\alpha}_{p0}$ is the average intrinsic molecular polarizability. In addition, in the case of liquids one can introduce a Clausius-Mossotti type of relation for the local gradient $\nabla\mathbf{E}$ to the macroscopic quadrupole moment density $\mathbf{Q}$ and a reaction field (similar to the relation between local field and average macroscopic polarization[14] $\mathbf{P}$). The local field is investigated in Refs. 12 and 13. We shall not attempt such a generalization in our study and in what follows we will assume that the equation of state (16) is valid for isotropic fluids, provided that $\mathbf{E}$ and $\nabla\mathbf{E}$ are not too large (in order Eq (12) to be applicable). For dense fluids, Eq (17) for $\alpha_Q$ will be invalid but it still must give the correct order of magnitude of the quadrupolarizability.

Using the values of the quadrupole moment of water from Ref. 37: $q_{xx} = +5.85\times10^{-40}$ Cm$^2$, $q_{yy} = -5.56\times10^{-40}$ Cm$^2$ and $q_{zz} = -0.29\times10^{-40}$ Cm$^2$ (a factor of 2/3 for the different definitions of $\mathbf{q}_0$ used here and in Ref. 37 must be accounted for), we can calculate the value $\alpha_Q = 1\times10^{-30}$ Fm from Eq (17). Both $\bar{\alpha}_{q0}$ and the Clausius-Mossotti effect increase $\alpha_Q$. For comparison, the experimental value for the polarizability of water is $\alpha_P = \varepsilon - \varepsilon_0 = 6.8\times10^{-10}$ F/m, which is about 3 times higher than the one calculated through the estimation $\alpha_P = Cp_0^2/3T$. By analogy, we can assume that $\alpha_Q$ is several times larger than the value following from Eq (17).

Let us now estimate the pressure and temperature derivatives of $\alpha_Q$. Assuming that the molecular quadrupolarizability $\alpha_q$ is independent on $p$, from Eq (17) it follows that

$$\frac{1}{\alpha_Q}\left(\frac{\partial\alpha_Q}{\partial p}\right)_T \approx \frac{1}{C}\left(\frac{\partial C}{\partial p}\right)_T = \beta_T , \qquad (22)$$

where $\beta_T$ is the isothermal compressibility. Since Eq (17) is approximate, the result Eq (22) also gives only an estimate of $\partial\alpha_Q/\partial p$. For the temperature derivative of $\alpha_Q$ (suitably made dimensionless by a factor of $T/\alpha_Q$), we use Eq (20) for the dependence of the molecular quadrupolarizability on temperature and the relation $\alpha_Q = C\alpha_q$ to obtain:

$$\frac{T}{\alpha_Q}\left(\frac{\partial\alpha_Q}{\partial T}\right)_p = \frac{T}{C}\left(\frac{\partial C}{\partial T}\right)_p + \frac{T}{\alpha_q}\left(\frac{\partial\alpha_q}{\partial T}\right)_p =$$
$$= -T\alpha_p^v - \frac{\mathbf{q}_0:\mathbf{q}_0/10T}{\bar{\alpha}_{q0}+\mathbf{q}_0:\mathbf{q}_0/10T} . \qquad (23)$$

Here $\alpha_p^v = -C^{-1}(\partial C/\partial T)_p$ is the coefficient of thermal expansion. For water[46], $T\alpha_p^v = 0.0763$. To estimate the second term, we assume that $\bar{\alpha}_{q0} \ll \mathbf{q}_0:\mathbf{q}_0/10T$ (water has high quadrupole moment $\mathbf{q}_0$ and it is a "hard" molecule of low intrinsic polarizability $\bar{\alpha}_{p0}$ and perhaps low $\bar{\alpha}_{q0}$). In this limit, the second term in Eq (23) is about –1, much larger in absolute value than $T\alpha_p^v$. Therefore, we can write approximately that

$$\frac{T}{\alpha_Q}\frac{\partial\alpha_Q}{\partial T} = -1 . \qquad (24)$$

## 3. Boundary conditions for the generalized Poisson equation

Within the quadrupolar approximation, the Coulomb-Ampere law (7) is of fourth order with respect to $\phi$ since upon substituting Eq (16) in Eq (7) one obtains:

$$\nabla\cdot\left[\varepsilon\mathbf{E} - \frac{1}{2}\nabla\cdot\alpha_Q\left(\nabla\mathbf{E} - \mathbf{U}\nabla\cdot\mathbf{E}/3\right)\right] = \rho(\phi) . \qquad (25)$$

In a homogeneous medium, this equation simplifies to

$$\nabla\cdot\mathbf{E} - \frac{\alpha_Q}{3\varepsilon}\nabla^2\nabla\cdot\mathbf{E} = -\nabla^2\phi + L_Q^2\nabla^4\phi = \frac{\rho(\phi)}{\varepsilon} . \qquad (26)$$

Here, we have introduced the *quadrupolar length* $L_Q$ defined with the relation:

$$L_Q^2 = \alpha_Q/3\varepsilon . \qquad (27)$$

From the estimation of $\alpha_Q$ in the end of the previous Section 2, we can say that $L_Q = (\alpha_Q/3\varepsilon)^{1/2} > 0.2$Å, perhaps several times larger. Eqs (26)-(27) are of the same form as those of Chitanvis[12], with the only difference that he obtained different numerical coefficient in Eq (27). We are mainly concerned with spherical symmetry in this study, where Eq (26) reads:

$$\frac{1}{r^2}\frac{\mathrm{d}r^2 E_r}{\mathrm{d}r} - \frac{L_Q^2}{r^2}\frac{\mathrm{d}}{\mathrm{d}r}r^2\frac{\mathrm{d}}{\mathrm{d}r}\frac{1}{r^2}\frac{\mathrm{d}r^2 E_r}{\mathrm{d}r} = \frac{\rho(r)}{\varepsilon} . \qquad (28)$$

We will need an explicit expression for $\mathbf{Q}$ and $\nabla\cdot\mathbf{Q}$; the gradient and the divergence of $\mathbf{E}$ in spherical coordinates are given by:

$$\nabla\mathbf{E} = \begin{pmatrix} \mathrm{d}E_r/\mathrm{d}r & 0 & 0 \\ 0 & E_r/r & 0 \\ 0 & 0 & E_r/r \end{pmatrix}; \quad \nabla\cdot\mathbf{E} = \frac{\mathrm{d}E_r}{\mathrm{d}r} + \frac{2E_r}{r} . \qquad (29)$$

Then, from Eq (16) one obtains

$$\mathbf{Q} = \frac{\alpha_Q}{3}\left(\frac{\mathrm{d}E_r}{\mathrm{d}r} - \frac{E_r}{r}\right)\begin{pmatrix} 2 & 0 & 0 \\ 0 & -1 & 0 \\ 0 & 0 & -1 \end{pmatrix},$$

$$\nabla \cdot \mathbf{Q} = \frac{2\alpha_Q}{3} \left( \frac{d^2 E_r}{dr^2} + \frac{2}{r} \frac{dE_r}{dr} - \frac{2 E_r}{r^2} \right) \mathbf{e}_r, \qquad (30)$$

where $\mathbf{e}_r$ is a unit vector, collinear with the radius-vector.

The boundary conditions of Eq (7) have been derived recently by Graham and Raab[39], using the singular distributions approach of Bedeaux et al.[47-48] in the case of a flat boundary surface of an anisotropic medium with arbitrary equation of state; alternative derivation, again for flat boundary, was given in Ref. 40. Following the approach of Graham and Raab, we will deduce here the boundary conditions of Eq (7) at a *spherical* surface dividing two isotropic phases. First, we write the singular distributions of $\varepsilon \mathbf{E}$, $\mathbf{Q}$ and $\rho$:

$$\varepsilon \mathbf{E} = \eta^+ \varepsilon^+ \mathbf{E}^+ + \eta^- \varepsilon^- \mathbf{E}^-;$$
$$\mathbf{Q} = \eta^+ \mathbf{Q}^+ + \eta^- \mathbf{Q}^-; \qquad (31)$$
$$\rho = \eta^+ \rho^+ + \eta^- \rho^- + \delta \rho^S.$$

Here, $X^+$ and $X^-$ denote the corresponding physical quantities for the phase situated at $r > R$ and $r < R$, respectively; $\rho^S$ is the surface charge density; the notations $\eta^\pm$ and $\delta$ stand for the Heaviside function $\eta$ and the Dirac $\delta$-function:

$$\eta^+ = \eta(r-R); \quad \eta^- = \eta(R-r); \quad \delta = \delta(r-R). \qquad (32)$$

To obtain the necessary boundary conditions, we insert Eqs (31) into Eq (7) and use the irreducibility of $\eta^\pm$, $\delta$ and its derivative $\delta_1 = d\delta/dr$. We need first to calculate $\nabla \cdot \varepsilon \mathbf{E}$ and $\nabla\nabla:\mathbf{Q}$, where $\varepsilon \mathbf{E}$ and $\mathbf{Q}$ are given by Eq (31):

$$\nabla \cdot \varepsilon \mathbf{E} = \eta^+ \nabla \cdot \varepsilon^+ \mathbf{E}^+ + \eta^- \nabla \cdot \varepsilon^- \mathbf{E}^- + \delta \left( \varepsilon^+ E_r^+ - \varepsilon^- E_r^- \right);$$

$$\nabla\nabla:\mathbf{Q} = \eta^+ \nabla\nabla:\mathbf{Q}^+ + \delta \mathbf{e}_r \cdot (\nabla \cdot \mathbf{Q}^+) + \delta \nabla \cdot (\mathbf{e}_r \cdot \mathbf{Q}^+) + \delta_1 Q_{rr}^+(r) +$$
$$+ \eta^- \nabla\nabla:\mathbf{Q}^- - \delta \mathbf{e}_r \cdot (\nabla \cdot \mathbf{Q}^-) - \delta \nabla \cdot (\mathbf{e}_r \cdot \mathbf{Q}^-) - \delta_1 Q_{rr}^-(r). \qquad (33)$$

Using Eqs (33), we can write Eq (7) in the form

$$\eta^+ \left[ \nabla \cdot \left( \varepsilon^+ \mathbf{E}^+ - \frac{1}{2} \nabla \cdot \mathbf{Q}^+ \right) - \rho^+ \right] +$$
$$+ \eta^- \left[ \nabla \cdot \left( \varepsilon^- \mathbf{E}^- - \frac{1}{2} \nabla \cdot \mathbf{Q}^- \right) - \rho^- \right] -$$
$$-\delta \left\{ \rho^S - \left[ \varepsilon^+ E_r^+ - \frac{1}{2} \mathbf{e}_r \cdot (\nabla \cdot \mathbf{Q}^+) - \frac{1}{2} \nabla \cdot (\mathbf{e}_r \cdot \mathbf{Q}^+) + \frac{1}{2} \frac{dQ_{rr}^+}{dr} \right] \right.$$
$$\left. + \left[ \varepsilon^- E_r^- - \frac{1}{2} \mathbf{e}_r \cdot (\nabla \cdot \mathbf{Q}^-) - \frac{1}{2} \nabla \cdot (\mathbf{e}_r \cdot \mathbf{Q}^-) + \frac{1}{2} \frac{dQ_{rr}^-}{dr} \right] \right\}_{r=R}$$
$$-\delta_1 \frac{1}{2} \left( Q_{rr}^+(R) - Q_{rr}^-(R) \right) = 0. \qquad (34)$$

For the derivation of Eqs (33)-(34), we have used the properties of the singular functions: $\nabla \eta^+ = \mathbf{e}_r \delta$; $\nabla \eta^- = -\mathbf{e}_r \delta$; $\nabla \delta = \mathbf{e}_r \delta_1$; $\delta_1 Q_{rr}(r) = \delta_1 Q_{rr}(R) - \delta dQ_{rr}/dr|_{r=R}$. Decomposition of Eq (34) yields, first, the bulk equations for the two phases (the coefficients of $\eta^\pm$ in Eq (34)):

$$\nabla \cdot \left( \varepsilon^\pm \mathbf{E}^\pm - \frac{1}{2} \nabla \cdot \mathbf{Q}^\pm \right) = \rho^\pm. \qquad (35)$$

Setting the factor multiplying $\delta$ in Eq (34) to 0, we obtain a generalization of the Gauss law for the quadrupolar media:

$$\left[ \varepsilon^+ E_r^+ - \frac{1}{2} \mathbf{e}_r \cdot (\nabla \cdot \mathbf{Q}^+) - \frac{1}{2} \nabla \cdot (\mathbf{e}_r \cdot \mathbf{Q}^+) + \frac{1}{2} \frac{dQ_{rr}^+}{dr} \right]_{r=R} -$$
$$- \left[ \varepsilon^- E_r^- - \frac{1}{2} \mathbf{e}_r \cdot (\nabla \cdot \mathbf{Q}^-) - \frac{1}{2} \nabla \cdot (\mathbf{e}_r \cdot \mathbf{Q}^-) + \frac{1}{2} \frac{dQ_{rr}^-}{dr} \right]_{r=R} = \rho^S. \qquad (36)$$

The last term of Eq (34), proportional to $\delta_1$, results in a new boundary condition, which balances the quadrupole moment densities on the two sides of the spherical surface:

$$Q_{rr}^+(R) - Q_{rr}^-(R) = 0. \qquad (37)$$

We now substitute Eqs (30) into Eqs (36) and (37) and obtain the explicit form of the boundary conditions. Eq (36) reads:

$$\left[ \varepsilon^+ E_r^+ - \frac{\alpha_Q^+}{3} \left( \frac{d^2 E_r^+}{dr^2} + \frac{2}{r} \frac{dE_r^+}{dr} - \frac{2 E_r^+}{r^2} \right) - \frac{\alpha_Q^+}{3} \frac{2}{r} \left( \frac{dE_r^+}{dr} - \frac{E_r^+}{r} \right) \right]_{r=R} -$$
$$- \left[ \varepsilon^- E_r^- - \frac{\alpha_Q^-}{3} \left( \frac{d^2 E_r^-}{dr^2} + \frac{2}{r} \frac{dE_r^-}{dr} - \frac{2 E_r^-}{r^2} \right) - \frac{\alpha_Q^-}{3} \frac{2}{r} \left( \frac{dE_r^-}{dr} - \frac{E_r^-}{r} \right) \right]_{r=R} =$$
$$= \rho^S. \qquad (38)$$

The explicit form of Eq (37) is:

$$\alpha_Q^+ \left( \frac{dE_r^+}{dr} - \frac{E_r^+}{r} \right)_{r=R} - \alpha_Q^- \left( \frac{dE_r^-}{dr} - \frac{E_r^-}{r} \right)_{r=R} = 0; \qquad (39)$$

Subtracting Eq (38) and Eq (39), we obtain the relation:

$$D_r^+ - D_r^- = \rho^S, \qquad (40)$$

which is formally equivalent to the classical Gauss law, but one must keep in mind that $\mathbf{D}$ involves higher derivatives of the field $\mathbf{E}$, cf. Eq (6).

## 4. Effect of the quadrupolarizability of a medium on the Born energies, partial molar volumes and entropies of dissolved ions

In this part of our study the general equation (28) of electrostatics in quadrupolar media at spherical symmetry and its boundary conditions (38)-(39) derived in the previous sections will be used to solve several basic electrostatic problems of high significance to the physical chemistry of electrolyte solutions.

### 4.1. Point charge in an insulator

We solve Eq (28) with $\rho = e_i \delta(\mathbf{r})$. The general solution of the equation is:

$$E = \frac{1}{r^2} \left[ k_1 + k_2 \left( 1 + \frac{r}{L_Q} \right) \exp\left( -\frac{r}{L_Q} \right) + k_3 \left( 1 - \frac{r}{L_Q} \right) \exp\left( \frac{r}{L_Q} \right) \right]. \qquad (41)$$

In order to determine the three integration constants $k_1$, $k_2$ and $k_3$, we need to impose three conditions on $E$. The first one is to require $E$ to tend to a finite value as $r \to \infty$ (this gives $k_3 = 0$). The second condition is that the asymptotic behavior of $E$ at $r \to \infty$ is unaffected by the presence of quadrupoles, that is, the field of a point charge at $r \to \infty$ tends to $q/4\pi\varepsilon r^2$. This condition yields $k_1 = q/4\pi\varepsilon$ (the same result can be obtained by the Gauss law). There is one final condition needed to determine $k_2$. Our assumption is to require that $E$ tends to something finite as $r \to 0$ i.e. there is no singularity of $E$ at $r \to 0$, which yields $k_2 = -k_1$. Eq (28) has thus a *finite* solution, which is:

$$E = \frac{e_i}{4\pi\varepsilon} \frac{1}{r^2} \left[ 1 - \left( 1 + \frac{r}{L_Q} \right) \exp\left( -\frac{r}{L_Q} \right) \right]. \qquad (42)$$

Integration of this result gives the following formula for the electrostatic potential:

$$\phi = \frac{e_i}{4\pi\varepsilon} \frac{1 - \exp(-r/L_Q)}{r}. \qquad (43)$$

The potential in $r = 0$ is also finite, and its value at $r = 0$ is $\phi_0 = e_i/4\pi\varepsilon L_Q$. The ion has, therefore, a finite energy:

$$u_{el} = e_i \phi_0 / 2 = e_i^2 / 8\pi\varepsilon L_Q. \qquad (44)$$

This is in marked contrast to the case of ion in vacuum where the potential is diverging as $1/r$ and the electrostatic self-energy of a point charge is infinite (Fig. 1). For a point charge in water at $T = 25°C$, if $L_Q = 2$ Å, we obtain $\phi_0 = 92$ mV and $u_{el} = 3.6 \times T$. Eq 2.8 of Chitanvis[12] has the same form as Eq (43) (but his relation between $L_Q$ and $\alpha_Q$ is different). Eq (43) can be compared also to Eq 2.48 of Jeon and Kim[13], who obtained a divergent potential since they used another constitutive relation for $\mathbf{Q}$ and implied different conditions on their solutions to determine the integration constants. In order to corroborate our non-classical choice for the finite field condition, a different approach will be presented in the following Section 4.2 to derive the same result (43), by placing the charge into a spherical cavity of radius $R$ (at $r = R$ the boundary conditions derived in Section 3 will be applied) and then taking the limit $R \to 0$ of the resulting potential.

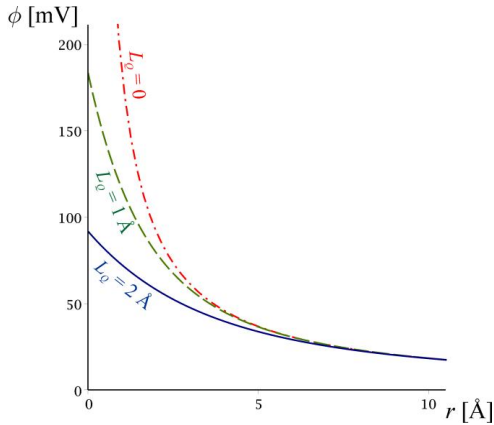

**Fig. 1.** Electrostatic potential $\phi$ of a point charge in a quadrupolar medium vs. the distance $r$ from the point charge in water, Eq (43), for various quadrupolar lengths $L_Q$. In a quadrupolar medium, the point charge has finite potential at $r = 0$. Solid line: $L_Q = 2$ Å; dashed line: $L_Q = 1$ Å; dash-dotted line: $L_Q = 0$ (the classical solution).

### 4.2. Ion of finite size in an insulator

There are various models of an ion of finite size in a solution, which yield the same expression for the Born energy[49,28,40]. The simplest model assumes that the ion is a point charge situated into a *cavity*, i.e., in an empty sphere of permittivity $\varepsilon_0$ and radius $R_{cav}$; the empty sphere is located in a medium of dielectric permittivity $\varepsilon$. This model neglects the detailed charge distribution in the ion and can be generalized in various ways[12,13,17,21,22,27-32,50]. Here, in order to keep the picture simple, we will hold on to the empty sphere model, only adding into account the quadrupolarizability $\alpha_Q$ of the medium. Similar problem (an entity of certain charge distribution placed into a cavity in a medium with intrinsic quadrupolarizability) was considered by Chitanvis[12] and Jeon and Kim[13] using different equation of state and a different set of boundary conditions.

We formulate the problem with the following equations:
**(i)** Inside the sphere (superscript "i"), at $r < R_{cav}$, there are no charges, bound or free, apart from the central ion of charge $e_i$:
$$\varepsilon_0 \nabla \cdot \mathbf{E}^i = e_i \delta(\mathbf{r}) . \tag{45}$$
**(ii)** Outside the sphere (no superscript), at $r > R_{cav}$, Eq (28) is valid with $\rho = 0$:
$$\frac{1}{r^2}\frac{dr^2 E_r}{dr} - \frac{L_Q^2}{r^2}\frac{d}{dr}r^2\frac{d}{dr}\frac{1}{r^2}\frac{dr^2 E_r}{dr} = 0 . \tag{46}$$
**(iii)** The boundary conditions at $r = R_{cav}$ are given by Eqs (38)-(39):
$$E_r - \frac{\varepsilon_0}{\varepsilon}E_r^i - L_Q^2\left(\frac{d^2 E_r}{dr^2} + \frac{2}{r}\frac{dE_r}{dr} - \frac{2E_r}{r^2}\right)\bigg|_{r=R_{cav}} = 0 ; \tag{47}$$

$$\left(\frac{dE_r^+}{dr} - \frac{E_r^+}{r}\right)_{r=R_{cav}} = 0 ; \tag{48}$$

The solution of Eqs (45)-(48) in terms of the electrostatic potential is:
$$\phi^i = \frac{e_i}{4\pi\varepsilon_0}\frac{1}{r} - \frac{e_i}{4\pi\varepsilon_0}\frac{1}{R_{cav}} + \frac{e_i}{4\pi\varepsilon}\frac{3L_Q + R_{cav}}{3L_Q^2 + 3L_Q R_{cav} + R_{cav}^2} \quad \text{at } r < R_{cav}; \tag{49}$$

$$\phi = \frac{e_i}{4\pi\varepsilon}\frac{1}{r}\left\{1 - \frac{3L_Q^2 \exp[-(r-R_{cav})/L_Q]}{3L_Q^2 + 3L_Q R_{cav} + R_{cav}^2}\right\} \quad \text{at } r > R_{cav}; \tag{50}$$

we used also the condition $\phi = \phi^i$ at $r = R_{cav}$. As $R_{cav} \to 0$, Eq (50) for $\phi$ simplifies to Eq (43) for a point charge, which justifies the assumption for finite $E$ and $\phi$ at $r \to 0$ which was used in Section 4.1 to derive Eq (43) – that is, the results obtained in Section 4.1 can be obtained alternatively by taking the limit from Eq (50) without making use of such a non-classical condition.

The self-energy $u_{el}$ of the ion is determined by the potential $\Delta\phi^i = \phi^i - e_i/4\pi\varepsilon_0 r$, created by the polarized medium and acting upon the ion. It is obtained from Eq (49) as
$$u_{el} = \frac{1}{2}e_i\Delta\phi^i = -\frac{e_i^2}{8\pi R_{cav}}\left(\frac{1}{\varepsilon_0} - \frac{1}{\varepsilon}\frac{3L_Q R_{cav} + R_{cav}^2}{3L_Q^2 + 3L_Q R_{cav} + R_{cav}^2}\right). \tag{51}$$

It can be compared to *Eq 3.1* of Chitanvis[12], who used instead of our Eq (48) a condition for continuity of $dE_r/dr$ at $r = R_{cav}$, without discussion. Jeon and Kim[13] used another condition – for non-oscillatory solution, also with no good justification. When $L_Q \to 0$, Eq (51) simplifies to the familiar expression for the Born energy[49,50]:
$$u_{Born} = -\frac{e_i^2}{8\pi R_{cav}}\left(\frac{1}{\varepsilon_0} - \frac{1}{\varepsilon}\right). \tag{52}$$

It follows from Eq (51) that the Born energy of small ions is more strongly affected from the quadrupolarizability of the medium.

The obtained result (51) for the effect of the quadrupolar length $L_Q$ on the self-energy $u_{el}$ is not easy to test directly due to the fact that, at least for water, the $1/\varepsilon_0$ term is by far larger than the second term in the brackets of Eq (51), which involves $L_Q$. To test Eq (51) we will eliminate the term $1/\varepsilon_0$ by differentiating $u_{el}$ either with respect to $p$ or to $T$ in Sections 4.4 and 0. Prior to this, we need first to discuss the relation of $R_{cav}$ to the crystallographic radius $R_i$ of the ion. Following Latimer, Pitzer and Slansky[50], we assume that
$$R_{cav,c} = R_i + L_c \quad \text{and} \quad R_{cav,a} = R_i + L_a \tag{53}$$
for cations and anions, respectively. Latimer, Pitzer and Slansky assumed that the lengths $R_i + L_c$ and $R_i + L_a$ are measures of the distance between an ion and the dipole closest to it. Rashin and Honig[51] argued that $R_i + L_c$ is, in fact, the covalent radius of the cation. The length $L_c$ is the same for all cations, and $L_a$ is the same for all anions; both are of the order of the water effective radius. Latimer, Pitzer and Slansky obtained $L_c = 0.85$ Å and $L_a = 0.1$ Å, using data for the free energies for hydration $\Delta\mu_{0i}$ of few monovalent ions ($L_c > L_a$ because $L_a$ must be about the distance of the anion to a proton and $L_c$ must be about the distance of a cation and an oxygen atom). We will determine these values more precisely in the following section, using a larger set of data for $\Delta\mu_{0i}$.

### 4.3. Ion free energy of hydration

The chemical potential of an ion in dilute aqueous solution is:
$$\mu_i = \mu_{0i} + T\ln C_i / C_0 . \tag{54}$$
The standard molarity-based chemical potential $\mu_{0i}$

(corresponding to a standard concentration $C_0$ = 1 M) reflects the state of a single ion in the solution, including the effect of the ion's field on the molecules of the solvent in the vicinity of the ion. For this energy one can write[2,17,22,52]:

$$\mu_{0i} = \mu_{intra}(T) + u_{el} + p\frac{4}{3}\pi R_i^3 g_V + T\ln C_0 . \quad (55)$$

The first term in Eq (55), $\mu_{intra}$, is related to the intramolecular state of the ion itself. For simple ions, it is assumed that this term is the same in gaseous state and in any solvent. The second term, $u_{el}$, stands for the electrostatic ion-water interaction; we will use for $u_{el}$ the generalized expression for the Born energy of the ion, Eq (51), involving the quadrupolar length $L_Q$. The third term, $pv_0$, is the mechanic work for introducing an ion of radius $R_i$ into a medium at pressure $p$; $g_V$ is a packing factor standing for the fact that the real volume $v_0$ occupied by the ion is not a sphere (of volume $4/3\,\pi R_i^3$) but rather a polyhedron. For example, if the packing of the solvent molecules around the ion is dodecahedral, then $g_V$ = 1.33; for a cubic packing, it is 1.6; for very large ions, $g_V$ is close to 1. Since for very small ions the term $pv_0$ is unimportant, and we are not going to analyze data for large ones due to the more complicated structure of their standard potential $\mu_{0i}$, we will assume that $g_V$ is approximately 1.33 for all ions studied below. The fourth term, $T\ln C_0$, originates from the choice of the standard state. Other contributions to Eq (55), such as the energy for cavity formation[52] and various specific interactions, are here neglected for simplicity. This makes Eq (55) inapplicable to large ions. Since the electrostatic effects we are investigating are significant for small ions only, this is not a drawback, but all data for molar hydration energies and partial molar volumes for ions larger than 3.2 Å will be neglected below. Hydrophobic effect is especially important for the partial molar entropy[53], therefore, only data for ions smaller than 2.3 Å will be taken into account. The full list of data-points is given in the Supplementary information C[†].

The standard molar free energy of hydration of an ion $\Delta\mu_{0i} = \mu_{0i} - \mu_{0i}^G$ is the energy for transfer of 1 mol ions from a hypothetical ideal gas at standard pressure $p_0$ to a hypothetical ideal 1M solution[54]. The expression for $\Delta\mu_{0i}$ follows from Eq (55) for $\mu_{0i}$ and an analogous expression for $\mu_{0i}^G$:

$$\Delta\mu_{0i} = T\ln\frac{TC_0}{p_0} + p\frac{4}{3}\pi R_i^3(g_V - 1)$$
$$-\frac{Z_i^2 e^2}{8\pi R_{cav}}\left(\frac{1}{\varepsilon_0} - \frac{1}{\varepsilon}\frac{3L_Q R_{cav} + R_{cav}^2}{3L_Q^2 + 3L_Q R_{cav} + R_{cav}^2}\right). \quad (56)$$

The first term stands for the choice of standard states in the aqueous solution (hypothetical ideal solution with concentration $C_0$ = 1M = 1000$\times N_A$ m$^{-3}$, $N_A$ – Avogadro's number) and in the gas state (ideal gas of ions at standard pressure $p_0$ = 101325 Pa). The second term stands for the mechanical work for introducing an ion of volume $4/3\,\pi R_i^3$ into the aqueous solution. This term is negligible compared to the other ones in Eq (56). The third term is the electrostatic energy for transferring an ion from a gas phase to an aqueous solution. The electrostatic energy in the gas (the $1/\varepsilon_0$ term in the brackets) is about 100 times larger than the respective energy in the solution (the term proportional to $1/\varepsilon$). Therefore, our expression (56) for $\Delta\mu_{0i}$ yields essentially the same results for the value of $\Delta\mu_{0i}$ as those of Latimer, Pitzer and Slansky who did not accounted for $L_Q$. Nevertheless, since we will use a more extended set of thermodynamic data for ions, we repeated their calculations by fitting the data for $\Delta\mu_{0i}$ taken from Refs. 54 and 52. Data-points for large ions as well as certain ions of high polarizability or dipole moments were neglected (cf. Supplementary information C[†] for the list).

The merit function is defined as:
$$\sigma_{\Delta\mu}^2(L_c, L_a, L_Q) =$$
$$= \frac{\sum_{Z=1}^{4}\sum_{i}\left[\Delta\mu_{0i,th}(L_c, L_a, L_Q; Z, R_i) - \Delta\mu_{0i,exp}(Z, R_i)\right]^2}{N - f}, \quad (57)$$

where $\Delta\mu_{0i,th}$ is the predicted value according to Eq (56) of an ion of valence Z and bare radius $R_i$; $\Delta\mu_{0i,exp}$ is the respective experimental value; N=86 is the number of data points and f is the number of free parameters used in the optimization procedure. This merit function is almost independent on $L_Q$. The results from the minimization of $\sigma_{\Delta\mu}$ with respect to $L_c$ and $L_a$ are given in Table 1 for various values of $L_Q$. They are in good agreement with the results of Latimer, Pitzer and Slansky, $L_c$ = 0.85 Å and $L_a$ = 0.1 Å, and are almost independent on $L_Q$. The comparison of Eq (56) with experimental data is illustrated in Fig. 2.

**Table 1.** Values of $L_c$ and $L_a$ obtained from the fit of the experimental data for the hydration energies of various ions with the theoretical expression, Eq (56), at three different values of $L_Q$.

| f | $L_c$ [Å] | $L_a$ [Å] | $L_Q$ [Å] | $\sigma_{\Delta\mu}$ [kJ/mol] |
|---|---|---|---|---|
| 2 | 0.84±0.06 | 0.15±0.14 | 2 | 152.0 |
| 2 | 0.84±0.06 | 0.15±0.14 | 1 | 151.9 |
| 2 | 0.83±0.06 | 0.14±0.14 | 0 | 151.9 |

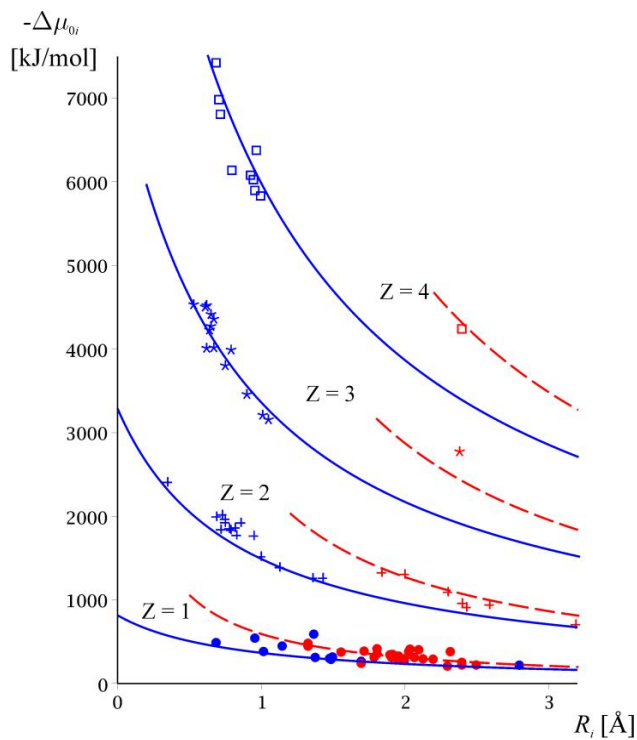

**Fig. 2.** Molar hydration energy $-\Delta\mu_{0i}$ [kJ/mol] vs. bare ion radii $R_i$ [Å]. Data for cations (blue) and anions (red) of various valence (circles – monovalent, crosses – divalent, stars – trivalent, squares – tetravalent). The lines are drawn according to the theoretical prediction Eq (56), with $L_c$ = 0.84 Å and $L_a$ = 0.15 Å, as obtained from the minimization of $\sigma_{\Delta\mu}$, Eq (57), at fixed value of $L_Q$ ($L_Q$ = 1Å).

Although Eq (56) is in satisfactory agreement with the experimental data, one must keep in mind it is an oversimplified model of an ion in a medium. The strongest assumption used in its derivation is that the continual model neglects the discrete nature of the solvent-ion interactions. The expression for the Born energy was corrected by many authors in order to take an explicit account for the discrete structure of matter (cf. *Chapter 5.7* of Ref. 17 for a summary). The homogeneity condition $\nabla \varepsilon = 0$ has been criticized e.g. by Abe[28]; the effect of dielectric saturation has been analyzed by Laidler and Pegis[26]. Nonlocal electrostatic theory was applied to the self-energy problem by Basilevsky and Parsons[31,32] (note that the presence of $\boldsymbol{Q} \propto \nabla \boldsymbol{E}$ in the definition (6) of the displacement field $\boldsymbol{D}$ makes the electrostatic problems in quadrupolar media nonlocal[13]). All these effects contribute to the value of $\mu_{0i}$, but these corrections will be neglected in the discussions below. In addition, the validity of Eq (16) for liquids is a hypothesis only. Therefore, the comparison of Eq (56) and its derivatives (Eqs (58) and (66) below) with the experimental data should be considered with caution.

### 4.4. Ion partial molecular volume

The models for the partial molecular volume $v_i$ of ions are reviewed in Ref. 55. We will consider only the values of $v_i$ at infinite dilutions. The partial molecular volume $v_i$ of the ion in aqueous solution is calculated by taking the derivative of $\mu_{0i} + T \ln C_i / C_0$, cf. Eqs (54)-(55), with respect to $p$:

$$v_i = \frac{\partial \mu_i}{\partial p} = \frac{4}{3}\pi R_i^3 g_V + T\beta_T + \frac{Z_i^2 e^2}{8\pi \varepsilon_0 R_{cav}^2}\frac{\partial R_{cav}}{\partial p} -$$
$$- \frac{Z_i^2 e^2}{8\pi \varepsilon^2} \frac{3L_Q + R_{cav}}{3L_Q^2 + 3L_Q R_{cav} + R_{cav}^2} \frac{\partial \varepsilon}{\partial p} -$$
$$- \frac{Z_i^2 e^2}{8\pi \varepsilon} \frac{3L_Q(3L_Q + 2R_{cav})}{(3L_Q^2 + 3L_Q R_{cav} + R_{cav}^2)^2} \frac{\partial L_Q}{\partial p}. \quad (58)$$

Here $\beta_T = -\partial \ln v_w / \partial p$ is the compressibility of water (the term $T\beta_T$ is relatively small and is usually neglected); $v_w$ is water's molar volume. In the third term proportional to $\partial R_{cav}/\partial p$, we neglected $1/\varepsilon$ in comparison with $1/\varepsilon_0$. If $L_Q = 0$, our expression (58) simplifies to the familiar formula[55] for $v_i$ following from the Born energy (52):

$$v_i = \frac{\partial \mu_i}{\partial p} = \frac{4}{3}\pi R_i^3 g_V + T\beta_T + \frac{Z_i^2 e^2}{8\pi \varepsilon_0 R_{cav}^2}\frac{\partial R_{cav}}{\partial p} - \frac{Z_i^2 e^2}{8\pi \varepsilon^2 R_{cav}}\frac{\partial \varepsilon}{\partial p}. \quad (59)$$

The expression (58) predicts the limiting partial molecular volume of an ion at infinite dilutions as a function of the ion crystallographic radius $R_i$.

While the hydration energy $\Delta\mu_{0i}$ is virtually independent of $L_Q$, the partial molar volume is sensitive to the value of $L_Q$, which allows us to use Eq (58) to determine $L_Q$ from the experimental data. We use the data for cations and anions of various valence assembled by Marcus[54], neglecting ions of complex structure and large $R_i$, cf. Supplementary information C[†]. The merit function of the optimization procedure is defined as:

$$\sigma_v^2\left(g_V, L_Q, \frac{\partial L_c}{\partial p}, \frac{\partial L_a}{\partial p}, \frac{\partial L_Q}{\partial p}\right) =$$
$$= \frac{\sum_{Z=1}^{4}\sum_i \left[v_{i,\text{th}}\left(Z, R_i; g_V, L_Q, \frac{\partial L_c}{\partial p}, \frac{\partial L_a}{\partial p}, \frac{\partial L_Q}{\partial p}\right) - v_{i,\exp}(Z, R_i)\right]^2}{N - f}; \quad (60)$$

the total number of data-points used in the optimization is N = 97.

First we need to estimate all parameters in Eq (58). The dependence $\varepsilon(p)$ was determined by the direct measurements[56] and allows the calculation of $\partial \varepsilon / \partial p$; we use the following value[55]:

$$\frac{1}{\varepsilon}\frac{\partial \varepsilon}{\partial p} = 4.76 \times 10^{-10} \text{ Pa}^{-1}. \quad (61)$$

The value of the compressibility is[55] $\beta_T = 4.57 \times 10^{-10}$ Pa$^{-1}$ and it corresponds to a partial molar volume $N_A T\beta_T = 1.13$ mL/mol. The values of $1/\varepsilon \times \partial\varepsilon/\partial p$ and $\beta_T$ are very close to each other since $\varepsilon$ is almost linear function of the water concentration – compare to Eq (22) for $\alpha_Q$. Using Eqs (61) and (22) we can estimate $\partial L_Q/\partial p$:

$$\frac{1}{L_Q}\frac{\partial L_Q}{\partial p} = \frac{1}{2\alpha_Q}\frac{\partial \alpha_Q}{\partial p} - \frac{1}{2\varepsilon}\frac{\partial \varepsilon}{\partial p} \approx -0.095 \times 10^{-10} \text{ Pa}^{-1}. \quad (62)$$

This value is quite small and therefore the last term in the brackets of Eq (58) plays little role for the partial molar volume $v_i$ of an ion in water and can even be neglected. For the geometry factor $g_V$ we can use the value 1.33, cf. the discussion following Eq (55). In all cases, we used the values $L_c = 0.84$ Å and $L_a = 0.15$ Å obtained in the previous section, cf. Table 1, when calculating $R_{cav} = R_i + L_{c,a}$. The value of $\partial R_{cav}/\partial p$ is in general different for cations and anions, $\partial R_{cav}/\partial p = \partial L_c/\partial p$ or $\partial L_a/\partial p$. We tested against the experimental data two simplifying assumptions regarding the two derivatives $\partial L_{a,c}/\partial p$ in order to decrease the number of free parameters. The first possibility investigated is that they are equal,

$$\partial L_c / \partial p = \partial L_a / \partial p. \quad (63)$$

The second one is that the quantity $1/L_{a,c} \times \partial L_{a,c}/\partial p$ is the same for both cations and anions:

$$\frac{1}{L_c}\frac{\partial L_c}{\partial p} = \frac{1}{L_a}\frac{\partial L_a}{\partial p}. \quad (64)$$

We tested both assumptions and Eq (64) was found to be in much better agreement with the experimental data, cf. Table 2.

We tested various combinations of fixed and free parameters for the optimization procedure (Table 2), in order to test the sensitivity of $\sigma_v$ to the parameters and the assumed approximations for $g_V$, $\partial R_{cav}/\partial p$, $\partial L_Q/\partial p$. In summary, the results are:

**(i)** $\sigma_v$ has a shallow minimum and the uncertainty of the values of the fitting parameters is high. This is illustrated in Fig. S1 in Supplementary material B[†].

**(ii)** The assumption Eq (64) yield lower dispersion than Eq (63) (compare rows [c]2 and [d]2 in Table 2). If both $\partial L_c/\partial p$ and $\partial L_c/\partial p$ are used as free parameters, we obtain values which agree within the uncertainty with Eq (64) (cf. rows [b]2, [g]3 and [i]4). The derivative $L_c^{-1}\partial L_c/\partial p$ is found to be negative and has a value of the order of $0.1 \times 10^{-10}$ Pa. This value can be compared to the pressure dependence of the distance $L_w$ between two water molecules; since $v_w \sim L_w^3$,

$$\frac{1}{v_w}\frac{\partial v_w}{\partial p} = \frac{1}{L_w^3}\frac{\partial L_w^3}{\partial p} = \frac{3}{L_w}\frac{\partial L_w}{\partial p}. \quad (65)$$

Therefore, $L_w^{-1}\partial L_w/\partial p = -\beta_T/3 = -1.5 \times 10^{-10}$ Pa. Thus $L_w^{-1}\partial L_w/\partial p$ is one order of magnitude larger than $L_c^{-1}\partial L_c/\partial p$, which suggests that the structure of the hydration shell of an ion is far more incompressible than the structure of water itself.

**(iii)** We tested whether the assumed value $g_V = 1.33$ of the packing factor gives good results by allowing $g_V$ to be a free parameter. This yielded a better dispersion and a slightly higher value: $g_V \sim 1.4$-$1.5$ (cf. rows [a]2, [f]3 and [i]4 in Table 2). This suggests that the packing of the hydration shell around the ion is less dense than dodecahedral ($g_V = 1.33$) but denser than cubic ($g_V = 1.6$).

**(iv)** We tested whether the approximate value $L_Q^{-1}\partial L_Q/\partial p = -0.095 \times 10^{-10}$ Pa$^{-1}$, (62), yields good results by considering it as a free parameter. Unfortunately, this returned almost the same dispersion and unrealistic values of $L_Q^{-1}\partial L_Q/\partial p$ and $L_Q$, since $\sigma_v$ is almost independent on this quantity. Therefore, we consider the results in rows [e]2 and [h]3 in Table 2 inadequate and use $L_Q^{-1}\partial L_Q/\partial p = -0.095 \times 10^{-10}$ Pa$^{-1}$.

**(v)** When one accounts for the effect of $L_Q$ on $v_i$, this yields only slightly lower dispersion (compare rows [a]2 and [f]3 or [b]2 and [g]3 in

Table 2). $L_Q$ affects the data for the smallest ions only (Li$^+$, Be$^{2+}$, Al$^{3+}$) and it explains why their partial molar volumes are more positive than the ones predicted from the classical model with $L_Q$ = 0. For example, the partial molar volume calculated for Al$^{3+}$ ($R_i$ = 0.53Å) with the parameters in row $^f$3 is -70 mL/mol, and if one sets $L_Q$ = 0, the result will be -85 mL/mol. The experimental value is[54] -59 mL/mol. The value of $L_Q$ obtained from the various variants of the optimization procedure varies between 1 and 2Å. From the estimation of $\alpha_Q$ in Section 2, we can predict that $L_Q$ is few times larger than 0.2Å. Still, a difference of one order of magnitude between the value of $L_Q$ estimated from Eqs (17) and (27) and the experimental one is unexpected. Nevertheless, the value of $L_Q$ will be confirmed with independent data for the partial molar entropy of various ions and data for the activity coefficient in the following two sections.

The comparison between Eq (58) and the experimental data is illustrated in Fig. 3 (parameters from row $^c$2).

**Table 2.** Results from the optimization of $\sigma_v$, Eq (60), with respect to various parameters. Blue fields indicate fixed values of the respective parameters.

| f | $L_Q$ [Å] | $\frac{1}{L_Q}\frac{\partial L_Q}{\partial p}$ [Pa$^{-1}$] ×10$^{10}$ | $\frac{1}{L_c}\frac{\partial L_c}{\partial p}$ [Pa$^{-1}$] ×10$^{12}$ | $\frac{1}{L_a}\frac{\partial L_a}{\partial p}$ [Pa$^{-1}$] ×10$^{12}$ | $g_V$ | $\sigma_v$ [mL/mol] |
|---|---|---|---|---|---|---|
| $^a$2 | 0 | 0 | -10.1 | Eq (64) | 1.43 | 12.4 |
| $^b$2 | 0 | 0 | -9.8 | -8.2 | 1.33 | 12.7 |
| $^c$2 | 1.1 | -0.095 | -11.3 | Eq (64) | 1.33 | 12.7 |
| $^d$2 | ~0 | -0.095 | -2.5 | Eq (63) | 1.33 | 17.9 |
| $^e$2 | 2.1 | 32.3 | 0 | 0 | 1.33 | 12.7 |
| $^f$3 | 2.1 | -0.095 | -13.1 | Eq (64) | 1.45 | 12.2 |
| $^g$3 | 2.0 | -0.095 | -12.7 | -10.2 | 1.33 | 12.6 |
| $^h$3 | -0.3 | -0.11 | -11.0 | Eq (64) | 1.33 | 12.7 |
| $^i$4 | 2.1 | -0.095 | -13.0 | -15.0 | 1.48 | 12.1 |

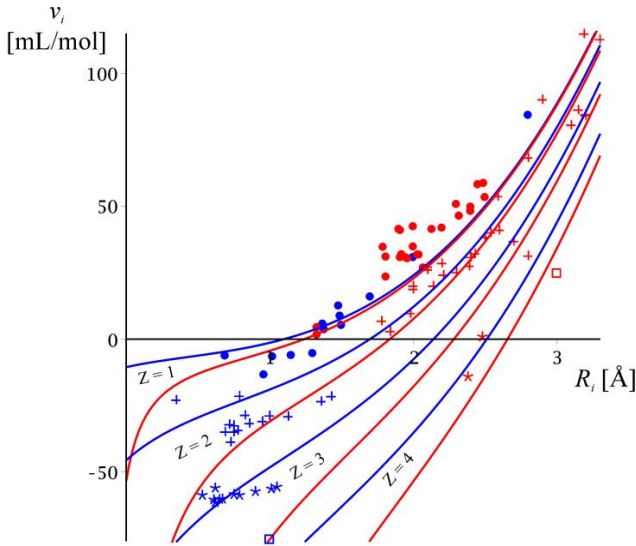

**Fig. 3.** Partial molar volumes $v_i$ [mL/mol] of ions of various valencies in infinitely diluted aqueous solutions as functions of the ionic crystallographic radii $R_i$. Solid circles: monovalent ions; crosses: divalent; stars: trivalent; boxes: tetravalent; blue and red – cations and anions. Data assembled by Marcus[54]. Lines: Eq (58) with Z = 1, 2, 3, 4 and $R_{cav} = R_i + L_c$ or $R_i + L_a$; the values for the parameters were obtained from the optimization of $\sigma_v$, Eq (60), with respect to two fitting parameters, $L_Q$ = 1.1 Å and $L_c^{-1}\partial L_c/\partial p$ = -11.3 10$^{-12}$ Pa$^{-1}$, for all 8 curves.

## 4.5. Standard entropy of hydration

The partial molar entropy $s_i$ of the ion in water can be calculated by taking minus the derivative of $\mu_{0i} + T\ln C_i/C_0$, cf. Eqs (54)-(55), with respect to $T$. The molar entropy *for hydration*, $\Delta s_i$, is calculated analogously as $-\partial(\Delta\mu_{0i} + T\ln C_i/C_0 - T\ln p/p_0)/\partial T$, cf. Eq (56) for $\Delta\mu_{0i}$. The *standard* molar entropy for hydration $\Delta s_{0i}$ is obtained2[54] by setting $p = p_0$ and $C_i = C_0$ in $\Delta s_i$. Using Eq (56), one finds the following expression for (the dimensionless) $\Delta s_{0i}$:

$$\Delta s_{0i} = -\ln\frac{TC_0}{p_0} - 1 + T\alpha_p^v - \frac{e_i^2}{8\pi R_{cav}^2}\frac{1}{\varepsilon_0}\frac{\partial R_{cav}}{\partial T} +$$
$$+ \frac{e_i^2}{8\pi\varepsilon^2}\frac{3L_Q + R_{cav}}{3L_Q^2 + 3L_Q R_{cav} + R_{cav}^2}\frac{\partial\varepsilon}{\partial T} +$$
$$+ \frac{e_i^2}{8\pi\varepsilon}\frac{3L_Q(3L_Q + 2R_{cav})}{(3L_Q^2 + 3L_Q R_{cav} + R_{cav}^2)^2}\frac{\partial L_Q}{\partial T}. \quad (66)$$

In Eq (66), all quantities except for $L_Q$, $\partial L_Q/\partial T$ and $\partial R_{cav}/\partial T = \partial L_{c,a}/\partial T$ are known.

The hydration entropy $\Delta s_{0i}$ has been measured with reasonable accuracy for a large number of ions[54]. The experimental data assembled by Marcus[54] can be used to obtain a second estimation of $L_Q$ from an independent set of data (besides the partial volumes), by comparing Eq (66) to them. To do so, we define the merit function:

$$\sigma_s^2\left(L_Q, \frac{\partial L_c}{\partial T}, \frac{\partial L_a}{\partial T}, \frac{\partial L_Q}{\partial T}\right) =$$
$$= \frac{\sum_{Z=1}^4 \sum_i \left[\Delta s_{0i,th}\left(Z, R_i; L_Q, \frac{\partial L_c}{\partial T}, \frac{\partial L_a}{\partial T}, \frac{\partial L_Q}{\partial T}\right) - \Delta s_{0i,exp}(Z, R_i)\right]^2}{N - f}. \quad (67)$$

Data for N = 68 ions of valence Z = 1÷4 are analyzed (cf. Supplementary material C).

The following values are used for the parameters in Eqs (66)-(67). For the temperature dependence of $\varepsilon$, we use the experimental data for $\varepsilon(T)$ from Refs. 2 and 46, which gives:

$$\frac{T}{\varepsilon}\frac{\partial\varepsilon}{\partial T} = -1.35. \quad (68)$$

For the coefficient of thermal expansion, we take[46] $T\alpha_p^v$ = 0.0763, corresponding to entropy of $k_B N_A \times 0.0763$ = 0.63 J/Kmol. We can also estimate $\partial L_Q/\partial T$ from Eqs (27), (24) and (68):

$$\frac{T}{L_Q}\frac{\partial L_Q}{\partial T} = \frac{T}{2\alpha_Q}\frac{\partial\alpha_Q}{\partial T} - \frac{T}{2\varepsilon}\frac{\partial\varepsilon}{\partial T} \approx 0.18. \quad (69)$$

Since this value is quite small, the term proportional to $\partial L_Q/\partial T$ will have insignificant contribution to the value of $\Delta s_{0i}$. The derivative $\partial R_{cav}/\partial T$ has different values for cations and anions, $\partial L_c/\partial T$ and $\partial L_a/\partial T$ respectively. In order to decrease the number of free parameters, we tested again two possible approximations, a first one that $\partial L_c/\partial T = \partial L_a/\partial T$ (which was found to be in disagreement with the experimental data), and a second one, that

$$\frac{1}{L_c}\frac{\partial L_c}{\partial T} = \frac{1}{L_a}\frac{\partial L_a}{\partial T}; \quad (70)$$

compare to the assumptions (63)-(64) for $\partial L_c/\partial p$ and $\partial L_a/\partial p$.

We again attempted various combinations of fixed and free parameters in the optimization procedure (Table 3), in order to analyze the sensitivity of $\sigma_s$ to $\partial R_{cav}/\partial T$, $\partial L_Q/\partial T$ and $L_Q$. The results are:

**(i)** $\sigma_s$ has a shallow minimum and does not allow for a very precise determination of the parameters involved. The dependence of $\sigma_s$ on $L_Q$ and $T/L_c \times \partial L_c/\partial T$ (cf. row $^d$2 of Table 3) is illustrated in Fig. S2 in Supplementary material B[†].

**(ii)** The experimental data agree well with Eq (70). If both $\partial L_c/\partial T$ and $\partial L_a/\partial T$ are used as free parameters, close values of $T/L_{c,a} \times \partial L_{c,a}/\partial T$ are obtained, cf. rows [c]2 and [f]3 in Table 3. The value of $T/L_c \times \partial L_c/\partial T$ is positive and has a value of the order of 0.02. This value can be compared to the temperature dependence of the distance $L_w$ between two water molecules; from the relation

$$\frac{1}{v_w}\frac{\partial v_w}{\partial T} = \frac{3}{L_w}\frac{\partial L_w}{\partial T} \quad (71)$$

we find that $T/L_w \times \partial L_w/\partial T = 3T\alpha_p^v = 0.23$. Similarly to the pressure dependence, cf. Eq (65), water expansion coefficient is higher by an order of magnitude compared to the respective dependence of $R_{cav}$ on $T$. This is another proof that the structure of water is more labile than the structure of the hydration shell of an ion.

**(iii)** The approximate value $T/L_Q \times \partial L_Q/\partial T = 0.18$ yield results which does not differ in comparison to the model with neglected $\partial L_Q/\partial T$ (rows [b]2 and [d]2). If $\partial L_Q/\partial T$ is left as a free parameter, the optimization procedure returns the same dispersion but unrealistic large negative values of $\partial L_Q/\partial T$ (row [e]3). Therefore, we consider the results in row [e]3 unrealistic and we use the value $T/L_Q \times \partial L_Q/\partial T = 0.18$. In fact, $T/L_Q \times \partial L_Q/\partial T$ can be safely neglected.

**(iv)** The effect of $L_Q$ on $\Delta s_{0i}$, and respectively, on the dispersion $\sigma_s$, is not very large. It mainly affects $\Delta s_{0i}$ of very small ions, by decreasing the absolute value of their entropies by 10-20%. For example, the entropy of Al$^{3+}$ calculated from Eq (66) with the parameters given in row [d]2 in Table 3 for $L_Q$ and $\partial L_c/\partial T$ is -597 J/Kmol, while with $L_Q = 0$ it is -678 J/Kmol (the experimental $\Delta s_{0i}$ of Al$^{3+}$ is[54] -538 J/Kmol). The results for the value of $L_Q$ does not depend strongly on the optimization procedure, and give values in the range $L_Q = 0.3-0.9$Å, somewhat smaller but of the same order as the prediction from the partial molar volume data. The value of $\alpha_Q$ corresponding to $L_Q = 0.80$Å (row [d]2) is $\alpha_Q = 3\varepsilon L_Q^2 = 13 \times 10^{-30}$ Fm, or about 13 times higher than the prediction from the ideal gas formula (17).

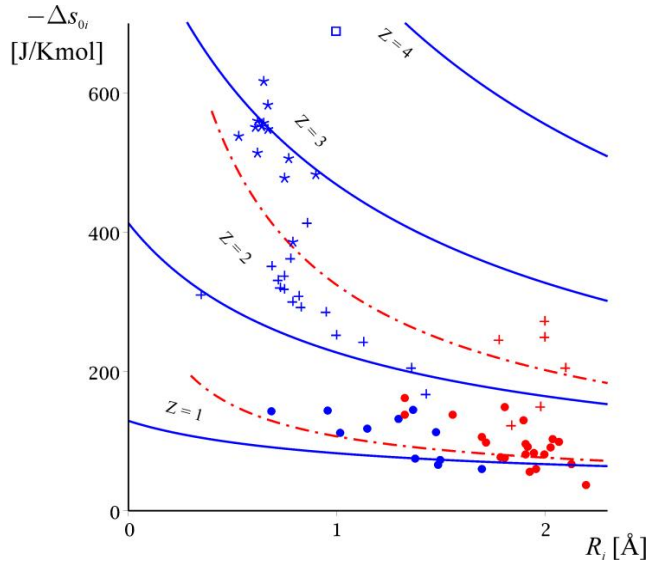

**Fig. 4.** Negative entropies of hydration $\Delta s_{0i}$ [J/Kmol] for mono, di, tri and tetravalent ions as functions of the crystallographic ionic radii $R_i$ [Å]. Data assembled by Marcus[54]. Lines: Eq (66) with two fitting parameters: $L_Q = 0.8$ Å and $T/L_c \times \partial L_c/\partial T = 0.025$, obtained from the minimization of $\sigma_s$, Eq (67). Solid circles: monovalent ions; crosses: divalent; stars: trivalent; boxes: tetravalent; blue and red – cations and anions.

**Table 3.** Results from the optimization of $\sigma_s$, Eq (67), with respect to various parameters. Blue fields indicate fixed values of the respective parameters.

| f | $L_Q$ [Å] | $\frac{T}{L_Q}\frac{\partial L_Q}{\partial T}$ | $\frac{T}{L_c}\frac{\partial L_c}{\partial T} \times 10^2$ | $\frac{T}{L_a}\frac{\partial L_a}{\partial T} \times 10^2$ | $\sigma_s$ [J/molK] |
|---|---|---|---|---|---|
| [a]1 | 0 | 0 | 2.0 | Eq (70) | 53.5 |
| [b]2 | 0.91 | 0 | 2.5 | Eq (70) | 53.4 |
| [c]2 | 0 | 0 | 2.0 | 2.4 | 53.2 |
| [d]2 | 0.80 | 0.18 | 2.5 | Eq (70) | 53.4 |
| [e]3 | 1.7 | -17.0 | -3.8 | Eq (70) | 53.3 |
| [f]3 | 0.34 | 0.18 | 2.2 | 2.5 | 53.2 |

## 5. Quadrupolarizability in the Debye-Hückel theory

The model of Debye-Hückel[1] of the electric double layer of a dissolved ion is a basic concept in electrolyte chemistry[2,17]. It has been corrected at least as many times as the Poisson-Boltzmann equation (4), but to our knowledge the corrections involved either the Boltzmann distribution or the homogeneity condition $\nabla \varepsilon = 0$. We are modifying the Poisson equation itself by accounting for the quadrupole term $\nabla\nabla:\mathbf{Q}$, Eq (7), and in this Section the effect of this term on the structure of the ionic atmosphere around an ion is investigated.

### 5.1. Point charge in conducting media

We solve Eq (28) with $\rho$ being given by the sum of a point charge and Boltzmann-distributed free charges, Eq (5):

$$\rho(\phi) = e_i\delta(\mathbf{r}) + \sum e_j C_j \exp(-e_j\phi/T). \quad (72)$$

Following Debye and Hückel[1], we expand the Boltzmann distribution in series at low potentials,

$$\rho(\phi) = e_i\delta(\mathbf{r}) - \varepsilon\phi/L_D^2; \quad (73)$$

here the Debye length $L_D$ is defined with the expression

$$L_D^2 = \varepsilon T / \sum e_j^2 C_j. \quad (74)$$

The solution of Eq (28) with $\rho$ given by Eq (73) is:

$$\phi = \frac{e_i}{4\pi\varepsilon}\frac{l_D^2 + l_Q^2}{l_D^2 - l_Q^2}\frac{\exp(-r/l_D) - \exp(-r/l_Q)}{r}, \quad (75)$$

where we used again the condition $\phi(0) < \infty$ to determine one integration constant, cf. Sections 4.1-4.2. In Eq (75), we introduced he characteristic lengths $l_D$ and $l_Q$ defined as

$$l_D = L_Q\left(\frac{1}{2} - \frac{1}{2}\sqrt{1 - 4L_Q^2/L_D^2}\right)^{-1/2},$$

$$l_Q = L_Q\left(\frac{1}{2} + \frac{1}{2}\sqrt{1 - 4L_Q^2/L_D^2}\right)^{-1/2}. \quad (76)$$

The inverse relations defining $L_D$ and $L_Q$ with $l_D$ and $l_Q$ are:

$$L_D^2 = l_D^2 + l_Q^2; \quad L_Q^{-2} = l_D^{-2} + l_Q^{-2}. \quad (77)$$

In dilute solutions, $L_D \gg L_Q$ and both $l_D$ and $l_Q$ are real; $l_D$ is about equal to $L_D$ and $l_Q$ is about equal to $L_Q$, which is the reason for the choice of indices. At a certain critical value of the Debye length, $L_D = 2L_Q$ (if $L_Q = 2$ Å, this correspond to ionic strength of 0.6 M), the lengths $l_D$ and $l_Q$ become equal (Fig. 5). At higher ionic strengths and smaller $L_D$, the lengths become complex conjugates, i.e., the potential (75) while diminishing with distance exhibits an oscillatory behavior. When $L_D < 2L_Q$, Eq (75) can be represented as:

$$\phi = \frac{e_i}{4\pi\varepsilon}\frac{l_{Re}^2 - l_{Im}^2}{l_{Re}l_{Im}r}\exp\left(-\frac{rl_{Re}}{l_{Re}^2 + l_{Im}^2}\right)\sin\left(\frac{rl_{Im}}{l_{Re}^2 + l_{Im}^2}\right), \quad (78)$$

where $l_{Re} = \mathrm{Re}\,l_D$ and $l_{Im} = \mathrm{Im}\,l_D$; it is easy to show that $l_{Im} =$

$(2L_D L_Q - L_D^2)^{1/2}/2$ and $l_{Re} = (2L_D L_Q + L_D^2)^{1/2}/2$. We will leave the deeper analysis for a future paper, since the oscillations of the potential are relatively unimportant for our current problem.

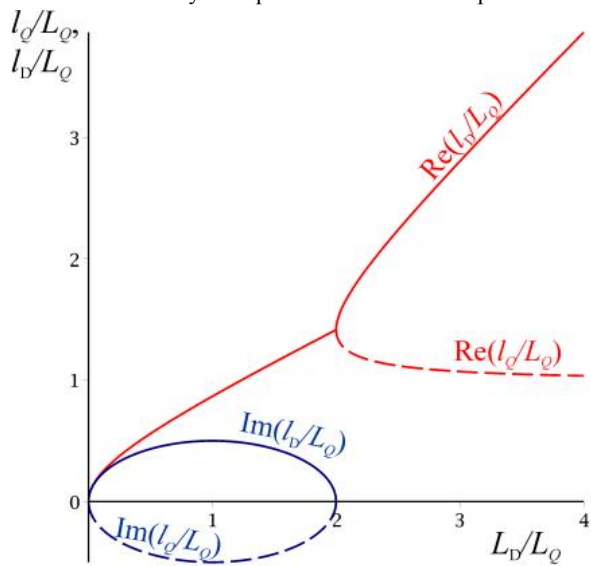

**Fig. 5.** Dimensionless characteristic lengths $l_D/L_Q$ and $l_Q/L_Q$ as functions of the dimensionless Debye length $L_D/L_Q$, Eqs (76). Red solid line: Re($l_D/L_Q$); red dashed line: Re($l_Q/L_Q$); blue lines: Im($l_Q/L_Q$) and Im($l_D/L_Q$).

The potential (75) is finite and its value at $r = 0$ is

$$\phi_0 = \frac{e_i}{4\pi\varepsilon l_Q l_D} \frac{l_D^2 + l_Q^2}{l_D + l_Q} =$$

$$= \frac{e_i}{4\pi\varepsilon L_Q} \frac{\sqrt{2}}{\left(1+\sqrt{1-4L_Q^2/L_D^2}\right)^{1/2} + \left(1-\sqrt{1-4L_Q^2/L_D^2}\right)^{1/2}}. \quad (79)$$

The respective energy $u_{el} = e_i\phi_0/2$ of the ion in the medium is

$$u_{el} = \frac{e_i^2}{8\pi\varepsilon l_Q l_D} \frac{l_D^2 + l_Q^2}{l_D + l_Q} =$$

$$= \frac{e_i^2}{8\pi\varepsilon L_Q} \frac{\sqrt{2}}{\left(1+\sqrt{1-4L_Q^2/L_D^2}\right)^{1/2} + \left(1-\sqrt{1-4L_Q^2/L_D^2}\right)^{1/2}}. \quad (80)$$

This expression can be expanded in series when $L_D \to \infty$ to obtain the limiting law for the energy $u_{el}$ in dilute solutions:

$$u_{el} \xrightarrow{L_D \to \infty} \frac{e_i^2}{8\pi\varepsilon L_Q} - \frac{e_i^2}{8\pi\varepsilon L_D} + \frac{3e_i^2 L_Q}{16\pi\varepsilon L_D^2} + ... \quad (81)$$

The first term in Eq (81) is the quadrupolar self-energy of a point charge, cf. Eq (44). The presence of the diffuse electric double layer decreases this energy with the Debye-Hückel energy $\mu_{DH}$:

$$\mu_{DH} \equiv T\ln\gamma_{DH} = -e_i^2/8\pi\varepsilon L_D. \quad (82)$$

Eq (82) is the well-known limiting Debye-Hückel law; $\gamma_{DH}$ is the activity coefficient of an ion in diluted electrolyte solution. The third term in Eq (81) represents the leading correction of the limiting Debye-Hückel law for a solution of point charges in a quadrupolar medium. It is positive, which means that the limiting law underestimates the activity coefficient $\gamma_{DH}$ at high concentration, which is indeed the case2. The correction for $L_Q$ is of the same order ($C_{el}^1$) as the correction for the finite ion size $R_i$, therefore, in order to compare the predicted effect of the quadrupolarizability on the activity coefficient with the experimental data, we need to generalize the *extended* Debye-Hückel model (the generalization of the limiting Debye-Hückel model for spherical instead of point charges).

## 5.2. Finite-size charge in conducting media

As was the case with the Born energy, the extended Debye-Hückel model can also be derived by various models of the ion and its cavity. The simplest one which yields the correct results assumes instead of Eq (73) the charge density:

$$\rho(\phi) = \begin{cases} e_i\delta(\mathbf{r}), & r < R, \\ -\varepsilon\phi/L_D^2, & r > R. \end{cases} \quad (83)$$

This equation reflects that the ions from the Debye atmosphere cannot approach the central ion to a distance smaller than $R$, which must be the sum of the two crystallographic radii $R_+$ and $R_-$ of the anion and the cation:

$$R = R_+ + R_-. \quad (84)$$

Eq (83) neglects the fact that two ions of the same sign can approach each other to distances $2R_+$ or $2R_-$ different from $R$; this is relatively unimportant since the repulsive electrostatic force decrease the co-ion concentration in the vicinity of the central ion to values close to 0. Also, the model assumes that $\varepsilon$ is the same inside the "cavity" and in the solution.

If Eq (83) is substituted into the Poisson equation (4), and this equation is solved, it will give for the activity coefficient the famous result, known as the extended Debye-Hückel theory1:

$$T\ln\gamma_{DH-R} = -\frac{e_i^2}{8\pi\varepsilon(L_D + R)}. \quad (85)$$

This equation has much wider applicability than the limiting law (82). However, instead of using Eq (84), the distance $R$ is always used as a fitting parameter and its value is generally higher than Eq (84) would predict – as pointed out by Israelashvilli[22], "*to obtain agreement with measured solubility and other thermodynamic data, it has been found necessary to "correct" the crystal lattice radii of ions by increasing them by 0.02 to 0.10 nm when the ions are in water*". The hypothesis that we investigate here is that this discrepancy reflects the neglected effect of the quadrupolar terms in Poisson equation.

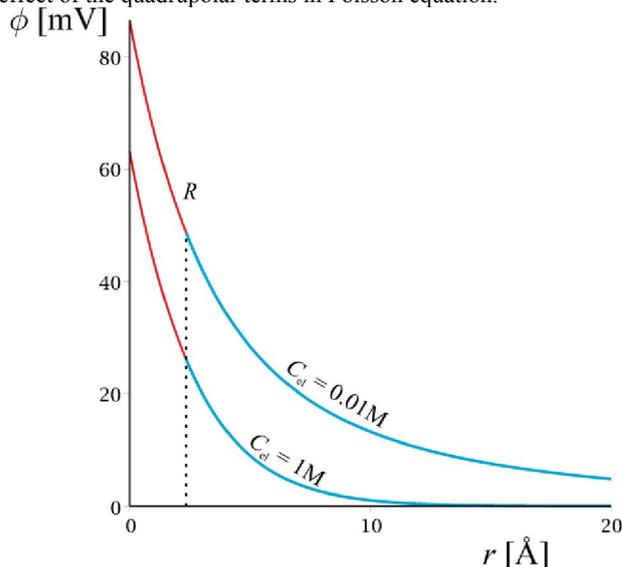

**Fig. 6.** Electrostatic potential $\phi$ of Na$^+$ ion in quadrupolar medium ($L_Q = 2$ Å) containing free charges (NaF of concentration $C_{el} = 0.01$ and 1 M). The minimal distance between Na$^+$ and F$^-$ is $R = 2.35$ Å (the sum of the crystallographic radii of Na$^+$ and F$^-$). The solution for $\phi$ at $C_{el} = 1$ M is oscillating-decaying and has shallow extrema (the first is $\phi = -0.0005$ mV at $r = 24.7$ Å).

To introduce $\alpha_Q$ in the extended Debye-Hückel model, we substitute Eq (83) into the generalization (28) of the Poisson equation of electrostatics (no jump of $\varepsilon$ or $\alpha_Q$ occurs at $r = R$).

The general solution of this problem is:

$$\phi = \begin{cases} A_0 + A_+ \dfrac{\exp\left(r\sqrt{l_D^{-2}+l_Q^{-2}}\right)-1}{r} - A_- \dfrac{\exp\left(-r\sqrt{l_D^{-2}+l_Q^{-2}}\right)-1}{r}, & r < R; \\ B_D \dfrac{\exp(-r/l_D)}{r} + B_Q \dfrac{\exp(-r/l_Q)}{r}, & r > R. \end{cases} \quad (86)$$

The five integration constants $A_0$, $A_+$, $A_-$, $B_D$ and $B_Q$ are determined by four conditions for continuity of $\phi$ and its derivatives (first, second and third) at $r = R$ and the electroneutrality condition:

$$e_i + \int_R^\infty \left(-\varepsilon\phi/L_D^2\right)4\pi r^2 dr = 0. \quad (87)$$

The solution of these 5 conditions is trivial but lengthy. The solution is given in Supplementary information D† (executable Maple 17 code). The final result for $\phi$ according to Eq (86) is illustrated in Fig. 6 at two concentrations of the electrolyte with $L_Q = 2$ Å and $R = 2.35$ Å.

From Eq (86) and the values of $A_0$, $A_+$, $A_-$, $B_D$ and $B_Q$, one can calculate the potential $\phi_0 = \phi(r=0)$ acting upon the central ion. It is simply related to the activity coefficient:

$$T\ln\gamma = \frac{e_i}{2}\left(\phi_0 - \frac{e_i}{4\pi\varepsilon L_Q}\right) = -\frac{e_i^2}{8\pi\varepsilon L_D} \times F(L_D, R, L_Q), \quad (88)$$

where the self-energy $-e_i^2/8\pi\varepsilon L_Q$ in the absence of other ions is subtracted, Eq (44). Here, we have introduced the function $F$:

$$F = -4l\left\{\left(l+d+\frac{1}{l\sqrt{l^2-1}}\right)\left[(l+1)e^d + (l-1)e^{-d}\right] + \frac{\sqrt{l^2-1}}{l}(l+d)\left[(l+1)e^d - (l-1)e^{-d}\right]\right\}^{-1}$$

$$+ \frac{l^2}{\sqrt{l^2-1}} \cdot \frac{\begin{aligned}&(l+1)\left(l+\sqrt{l^2-1}\right)e^{2d} + \\ &+(l-1)\left[l(2l+2d+1)-2\sqrt{l^2-1}(l+d)-\dfrac{l^2-3}{\sqrt{l^2-1}}\right]\end{aligned}}{\begin{aligned}&(l+1)\left[(l+d)\left(l+\sqrt{l^2-1}\right)+\dfrac{1}{\sqrt{l^2-1}}\right]e^{2d} \\ &+(l-1)\left[(l+d)\left(l-\sqrt{l^2-1}\right)+\dfrac{1}{\sqrt{l^2-1}}\right]\end{aligned}}; \quad (89)$$

where the dimensionless lengths $l$ and $d$ are defined as $l = l_D/L_Q$ and $d = R/L_Q$. In diluted solutions, Eq (88) can be expanded in series and the result is:

$$T\ln\gamma \xrightarrow{L_D \to \infty} -\frac{e_i^2}{8\pi\varepsilon L_D}\left[1 - \frac{2R + 4L_Q\exp(-R/L_Q) - L_Q\exp(-2R/L_Q)}{2L_D}\right], \quad (90)$$

which can be used at low electrolyte concentration. However, in the most interesting for our consideration region of concentrations, $C_{el} \sim 1M$, where the effect of $L_Q$ is most important, Eqs (88)-(89) must be used without simplifications.

The comparison of Eq (88) to the experimental data for the activity coefficient $\gamma$ allows the determination of the single unknown parameter in it, the quadrupolar length $L_Q$, provided that our model is suitable for the electrolyte under investigation. The effect of $L_Q$ is commeasurable with several other effects, e.g., the specific dispersion ion-ion interactions, correlation effects etc.[2,17]. In the case of small ions, direct electrostatic interaction will prevail over these effects. NaF is especially suitable for the comparison with Eq (88) since F[-] is the smallest anion, it has low polarizability and minimizes the non-electrostatic terms that are not included in the Boltzmann distribution (5) used in our derivation. Besides, Na[+] and F[-] have similar radii, which minimizes the error from the approximation that the double layer starts from $R_+ + R_-$ (instead of using 3 lengths, $R_+ + R_-$ for counter-ions and $2R_+$ or $2R_-$ for co-ions). The experimental data for the mean activity coefficient $\gamma_\pm$ of NaF is well-described by the formula[2,57]:

$$\lg\gamma_\pm = -\frac{A_D\sqrt{C_m}}{1+B\sqrt{C_m}} + \beta C_m, \quad (91)$$

where $A_D$ is the Debye-Hückel coefficient (-0.5108 kg$^{1/2}$/mol$^{1/2}$ at 25°), and the semi-empirical parameters $B$ and $b$ have the values[57] $B = 1.28$ kg$^{1/2}$/mol$^{1/2}$ and $\beta = -0.018$ kg/mol. Eq (91) is valid up to molal concentration $C_m = 1$ mol/kg. Since the partial molar volume of NaF is very small[54], the relation between molarity and molality is simply $C_{el}$ [mol/L] $= C_m\rho_m$, where $\rho_m = 0.997$ kg/L.

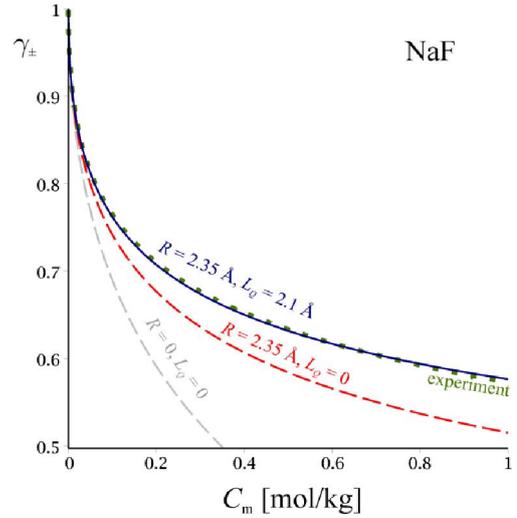

**Fig. 7.** Mean ion activity coefficient $\gamma_\pm$ vs. molality $C_m$. Comparison between the experimental interpolation formula Eq (91) (green dotted line) of the NaF data[57] and the generalized Debye-Hückel model which takes into account quadrupolarizability, Eqs (88)-(89), with $R = R_{Na} + R_F = 2.35$ Å following from the ionic crystallographic radii[54] and quadrupolar length $L_Q = 2.1$ Å obtained as a fitting parameter. For comparison, the limiting Debye-Hückel model ($R = L_Q = 0$), Eq (82), and the extended Debye-Hückel equation, Eq (85), with $R = 2.35$ Å are given.

The comparison between the experimental dependence (91) for NaF and Eqs (88)-(89) is shown in Fig. 7. To determine the best value of $L_Q$, we optimized the merit function

$$\sigma_\gamma(L_Q) = \int_0^{1\text{ mol/kg}} \left|\ln\gamma(C_m; L_Q) - \ln\gamma_\pm(C_m)\right| dC_m, \quad (92)$$

with $\gamma_\pm$ being the experimental mean activity coefficient given by Eq (91) and $\gamma$ is the theoretical activity coefficient from Eqs (88)-(89). The best value of $L_Q$ is 2.11±0.06 Å, in good agreement with the values found from the partial volume data and larger than those following from the entropy data. The agreement between theory and experiment is excellent (Fig. 7). For comparison, the extended Debye-Hückel model is also shown in Fig. 7, with $R = 2.35$ Å as predicted by Eq (84) (instead of using it as a fitting parameter). The limiting Debye-Hückel law ($L_Q = R = 0$) is also given for comparison. Note that the effect of $L_Q$ is quite significant – it is of the same order as the effect of $R$.

# 6. Conclusions

Our work investigates the effects of the quadrupole moment of the molecules in a medium on the properties of charged particles dissolved in this medium, using a macroscopic approach based on the quadrupolar Coulomb-Ampere law (26), generalizing the classical Poisson equation of electrostatics.

(i) We derived a new equation of state, Eq (16), relating the macroscopic density of quadrupole moment $Q$ and the field gradient $\nabla E$ in gas of quadrupoles. The tensor $Q$ has zero trace, unlike the one used in Refs. 12 and 13. Our constitutive relation involves a single scalar coefficient, the quadrupolarizability $\alpha_Q$, which was estimated to be $\alpha_Q = 1 \times 10^{-30}$ Fm or few times larger.

(ii) We derived the boundary conditions needed for the fourth-order quadrupolar Coulomb-Ampere law (26) at a spherical surface between two media of different dielectric permittivity $\varepsilon$ and quadrupolarizability $\alpha_Q$, Eqs. (38)-(39).

(iii) The potential of a point charge in quadrupolar medium is finite even at $r = 0$, cf. Eq (43). This unexpected result was obtained previously by Chitanvis[12] with another constitutive relation for $Q$.

(iv) The classical model for a dissolved ion as a charge in a cavity was generalized for the case of quadrupolar medium. It was shown that the quadrupolarizability of water affects significantly the thermodynamic properties – partial molar volume $v_i$ and entropy for hydration $\Delta s_{0i}$ – of small ions in aqueous solution. From this effect and the experimental thermodynamic data for $v_i$ and $\Delta s_{0i}$ from Ref. 54, the value of the quadrupolar length, $L_Q = (\alpha_Q/3\varepsilon)^{1/2}$ = 1-2 Å, could be estimated.

(v) The Debye-Hückel model for the diffuse ionic atmosphere of an ion was generalized by including the quadrupolarizability of the medium in it. Comparison with data for the activity coefficient $\gamma_\pm$ of NaF allowed independent determination of the value of $L_Q$, which yielded again $L_Q \sim 2$Å. The minimal distance of approach $R$ between ions in the extended Debye-Hückel model must not be corrected from the expected value, $R = R_+ + R_-$, where $R_+$ and $R_-$ are the crystallographic radii.

(vi) The order of magnitude of $\alpha_Q$ and $L_Q$ obtained from these 3 sets of experimental data ($v_i$, $\Delta s_{0i}$ and $\gamma_\pm$) compares well with the order predicted by other authors[12,13].

(vii) The pressure and temperature derivatives of $\alpha_Q$ were estimated theoretically, cf. Eqs (22)-(23). The estimated values of $\partial \alpha_Q/\partial p$ and $\partial \alpha_Q/\partial T$ show that the effect from these derivatives on the partial molar volume and entropy, Eqs (58) and (66), of the dissolved ion is negligible. The pressure and temperature derivatives of the radius of the cavity $R_{cav}$ around an ion were estimated from experimental data. From their values it can be concluded that the structure of the hydration shell of an ion is about 10 times stiffer than the structure of water.

Although the results obtained here are encouraging, one must not forget that our approach uses some strong approximations. First, the constitutive relation Eq (16) is strictly valid for diluted ideal gas only. The assumption that the equation of state keeps the same form in dense liquid needs additional justification. Also, our model for the dissolved ion (point charge in a cavity) is clearly an oversimplification, as discussed in Section 4.2 and 5.2. Nevertheless, we obtain self-consistent results and we have enough proof to assert that quadrupolarizability has measurable effect on many thermodynamic characteristics of the dissolved ions ($v_i$, $\Delta s_{0i}$ and $\gamma_\pm$).

The results obtained here for the equation of state for $Q$, the boundary condition for the generalized Maxwell equations of electrostatics and the value of the quadrupolarizability $\alpha_Q$ of water will be used in the following study of this series for the analysis of several important problems of colloid science.

# Acknowledgements

This work was funded by the Bulgarian National Science Fund Grant DDVU 02/12 from 2010. R. Slavchov is grateful to the FP7 project BeyondEverest. Consultations with Prof. Alia Tadjer are gratefully acknowledged.

# Notes and references